\documentclass[
 twocolumn,
superscriptaddress,
 amsmath,amssymb,
 aps,
prb,
raggedfooter,
]{revtex4-2}

\usepackage{graphicx}
\usepackage{dcolumn}
\usepackage{bm}
\usepackage[hidelinks]{hyperref}
\usepackage{tabularx}
\usepackage{flushend}
\usepackage{amsmath}
\usepackage{makecell}

\hypersetup{
  colorlinks   = true, 
  urlcolor     = blue, 
  linkcolor    = blue, 
  citecolor   = blue 
}

\begin{document}

\preprint{APS/123-QED}

\title{Selective coupling of high-order phonons in La$_{2-x}$Sr$_x$CuO$_4$}

\author{Ke-Jun Xu}
\thanks{equal contribution}
\affiliation{%
Material Sciences Division, Lawrence Berkeley National Laboratory, Berkeley, California 94720, USA
}
\affiliation{%
Department of Physics, University of California, Berkeley, California 94720, USA
}

\author{Paulina Majchrzak}
\thanks{equal contribution}
\affiliation{%
Stanford Institute for Materials and Energy Sciences, SLAC National Accelerator Laboratory, 2575 Sand Hill Road, Menlo Park, CA 94025, USA
}
\affiliation{%
Geballe Laboratory for Advanced Materials, Stanford University, Stanford, California 94305, USA 
}

\author{Nathan Giles-Donovan}
\thanks{equal contribution}
\affiliation{%
Department of Physics, University of California, Berkeley, California 94720, USA
}
\affiliation{%
Material Sciences Division, Lawrence Berkeley National Laboratory, Berkeley, California 94720, USA
}

\author{Sangheon Kim}
\affiliation{%
Department of Physics and Astronomy, Seoul National University, Seoul 08826, Republic of Korea
}

\author{Jaewon Choi}
\affiliation{%
Department of Physics and Astronomy, Seoul National University, Seoul 08826, Republic of Korea
}

\author{Jun Okamoto}
\affiliation{%
National Synchrotron Radiation Research Center, Hsinchu 30076, Taiwan
}

\author{Ganesha Channagowdra}
\affiliation{%
National Synchrotron Radiation Research Center, Hsinchu 30076, Taiwan
}

\author{Hsiao-Yu Huang}
\affiliation{%
National Synchrotron Radiation Research Center, Hsinchu 30076, Taiwan
}

\author{Di-Jing Huang}
\affiliation{%
National Synchrotron Radiation Research Center, Hsinchu 30076, Taiwan
}

\author{Seiki Komiya}
\affiliation{%
Central Research Institute of Electric Power Industry, Yokosuka 240-0196, Japan
}

\author{Takao Sasagawa}
\affiliation{%
Materials and Structures Laboratory, Institute of Science Tokyo, Yokohama, Kanagawa 226-8501, Japan
}

\author{Shuichi Wakimoto}
\affiliation{%
Japan Atomic Energy Agency, Tokai, Ibaraki 319-1195, Japan
}

\author{Wei-Sheng Lee}
\affiliation{%
Stanford Institute for Materials and Energy Sciences, SLAC National Accelerator Laboratory, 2575 Sand Hill Road, Menlo Park, CA 94025, USA
}

\author{Zhi-Xun Shen}
\affiliation{%
Stanford Institute for Materials and Energy Sciences, SLAC National Accelerator Laboratory, 2575 Sand Hill Road, Menlo Park, CA 94025, USA
}
\affiliation{%
Geballe Laboratory for Advanced Materials, Stanford University, Stanford, California 94305, USA 
}
\affiliation{%
Department of Applied Physics, Stanford University, Stanford, California 94305, USA
}
\affiliation{%
Department of Physics, Stanford University, Stanford, California 94305, USA
}

\author{Thomas P. Devereaux}
\affiliation{%
Stanford Institute for Materials and Energy Sciences, SLAC National Accelerator Laboratory, 2575 Sand Hill Road, Menlo Park, CA 94025, USA
}
\affiliation{%
Geballe Laboratory for Advanced Materials, Stanford University, Stanford, California 94305, USA 
}
\affiliation{%
Department of Materials Science and Engineering, Stanford University, Stanford, California 94305, USA 
}

\author{Dung-Hai Lee}
\email{dunghai@berkeley.edu}
\affiliation{%
Department of Physics, University of California, Berkeley, California 94720, USA
}
\affiliation{%
Material Sciences Division, Lawrence Berkeley National Laboratory, Berkeley, California 94720, USA
}

\author{Robert J. Birgeneau}
\email{robertjb@berkeley.edu}
\affiliation{%
Department of Physics, University of California, Berkeley, California 94720, USA
}
\affiliation{%
Material Sciences Division, Lawrence Berkeley National Laboratory, Berkeley, California 94720, USA
}

\date{\today}

\begin{abstract}
In crystals and molecules with strong coupling between the charge carriers and atomic vibrations, high order harmonics of the vibrational excitations can be observed with intensity following the Franck-Condon envelope. Here, we uncover a new regime of electron-phonon coupling in La$_{2-x}$Sr$_x$CuO$_4$ using resonant inelastic X-ray scattering. In the undoped compound, we find sharp peaks at approximately 85$\pm$5 meV, 180$\pm$7 meV, and 330$\pm$5 meV with no observable dispersion, and an absence of appreciable intensity from other modes observed in Raman spectroscopy in this energy range. The 180 meV and 330 meV excitations have energies approximately consistent with two-phonon and four-phonon in-plane Cu-O bond stretching mode, suggesting that the resonantly excited valence electrons strongly couple to this mode. These excitations exhibit significant doping dependence, eventually becoming unresolvable at 10\% hole doping, where the low energy charge excitations are dominated by dispersive plasmons. The observation of selective coupling to even-ordered phonons and stark contrast with the Raman spectra indicate anomalous electron-phonon coupling beyond the Franck-Condon picture. One intriguing possibility is the existence of locally paired quasiparticles in insulating cuprates.


\end{abstract}

\maketitle


Phonons are fundamental quanta of lattice vibrations that exist in all crystalline materials. By themselves, phonons are transporters of thermal and acoustic energy~\cite{nli1}. In metals and systems on the verge of itineracy, phonons can also couple to valence electrons and greatly influence the electronic structure and many physical properties~\cite{giustino1}. In most metals, electron-phonon scattering produces distinct temperature-dependent resistivity forms at high temperatures and low temperatures~\cite{bloch1, gruneisen1}, and act as the source of the attractive pairing interactions in conventional superconductors~\cite{bardeen1}. In strongly coupled cases, electron-phonon coupling can result in significant valence band renormalization~\cite{lanzara1, cuk1, devereaux1, anzai1}.

\begin{figure}
\includegraphics[width=0.4\textwidth]{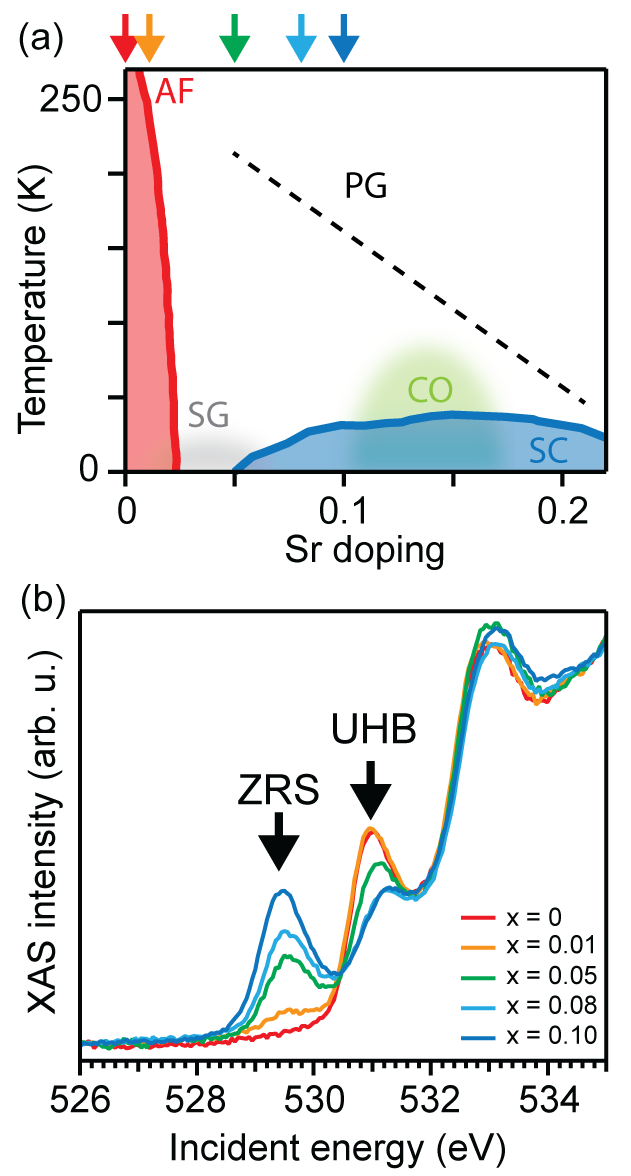}
\caption{X-ray absorption spectroscopy at the O K-edge for LSCO. (a) Temperature-doping phase diagram of LSCO. Blue region indicates superconductivity (SC), red region indicates long-ranged antiferromagnetism (AF), grey region indicates spin glass (SG), green region indicates charge order (CO). Black dashed line and grey data points indicates the approximate boundary of the pseudogap (PG) phase. Arrows at the top of the panel indicates the dopings used for the RIXS measurements. (b) X-ray absorption spectra at the O K-edge for the dopings indicated by the colored labels. ZRS indicates the Zhang-Rice singlet states, and UHB indicates the upper Hubbard band states. The measurements were taken with the $a-b$ plane surface facing the incoming X-ray beam. The incident photons have $\sigma$ polarization with an electric field in the vertical direction (perpendicular to the scattering plane) and along the Cu-O bond direction. Measurement temperature was 30 K.}
\end{figure}

Coupling between phonons and electrons has been extensively studied with single-particle probes, due to the variety of experimental tools available and their high energy resolution, including angle-resolved photoemission spectroscopy (ARPES)~\cite{sobota1} and scanning tunneling spectroscopy (STS)~\cite{you1}. In the cuprates, ARPES has shown that specific oxygen phonon modes can modify the electronic structure significantly with momentum selectivity~\cite{cuk1, devereaux1}, and are thought to play a synergistic role with spin fluctuations for enhancing superconductivity~\cite{tallon1, jlee1, johnston1, he1,guo1}.

\begin{figure*}[t]
\includegraphics[width=1\textwidth]{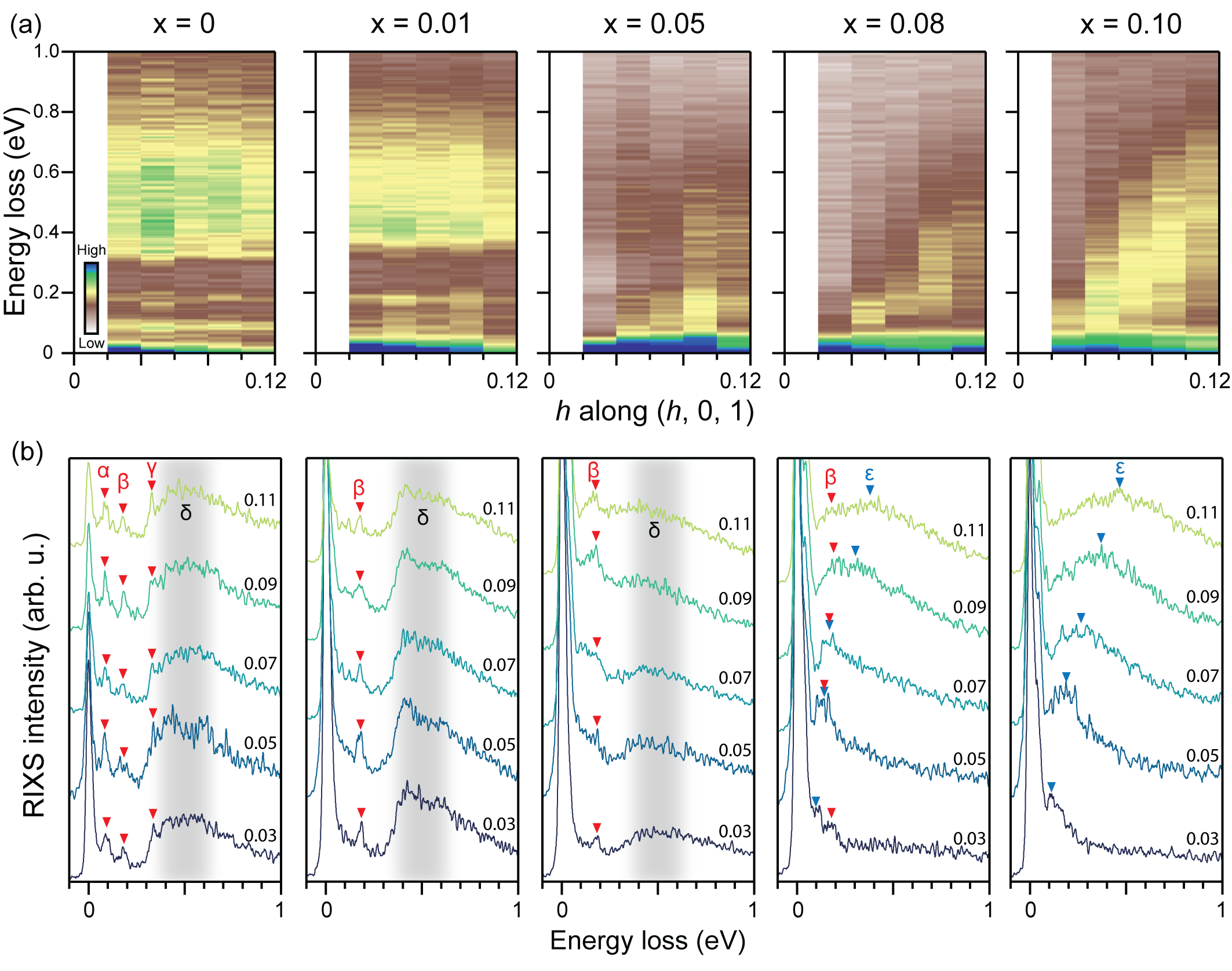}
\caption{Doping and momentum dependences of low energy excitations in LSCO. (a) Color image of the RIXS spectra as a function of energy loss and in-plane momentum $h$ along ($h$, 0, 1) for dopings indicated above each panel. (b) RIXS energy distribution curves for respective dopings as in (a). Numbers next to each curve indicate the in-plane momentum $h$ along ($h$, 0, 1). Greek letters are assigned for each distinct feature. $\alpha$, $\beta$, and $\gamma$ peaks (red markers) highlight the sharp low energy lattice excitations. The $\delta$ feature (grey shaded region) highlights the broad excitation around 400 to 600 meV. $\epsilon$ peaks (blue markers) highlight the dispersive excitation in x = 0.08 and x = 0.1 samples. In the x = 0.08 sample, it is not possible to obtain stable fits for the momentum range where the $\beta$ feature appear to cross the $\epsilon$ feature, and the peak is marked with a double-colored marker. RIXS measurements were performed with $\sigma$-polarized photons (vertical E field along the Cu-O bond direction) at a temperature of 30 K. The markers here are guides to the eye, and the fittings for the features of interest are shown elsewhere.} 
\end{figure*}

\begin{figure}[t]
\includegraphics[width=0.4\textwidth]{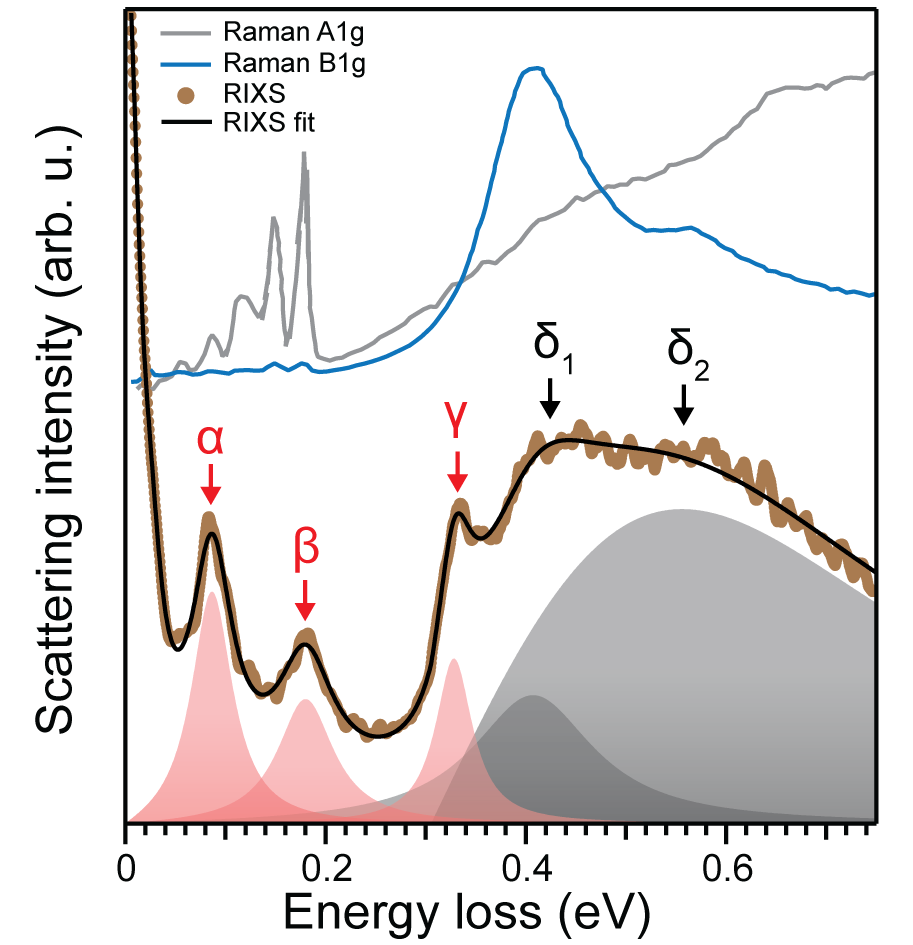}
\caption{Comparison of momentum-integrated RIXS spectrum with Raman spectra. Brown dots are the raw RIXS energy distribution curve for integrated momentum between $h$ = 0.03 to $h$ = 0.11 along ($h$, 0 , 1). Black line is the overall fit, comprised of an elastic line (not shown), a low energy antilorentzian for capturing the low energy phonons that are not individually resolved (not shown), four antilorentzians for the $\alpha$, $\beta$, $\gamma$, and $\delta_1$, and another antilorentzian where the zero is not fixed for the $\delta_2$ feature. The grey curve is the Raman spectrum in the $A_{1g}$ channel and the blue curve is the Raman spectrum in the $B_{1g}$ channel, both from ref~\cite{chelwani1}.}
\end{figure}

On the other hand, understanding quasiparticle transport in the strong electron-phonon coupling regime requires measurement of the two-particle correlation function, generally probed with scattering methods. The phonon spectrum can be probed with a number of techniques that are sensitive to the ions (inelastic neutron scattering, or INS)~\cite{shirane1}, core electrons (inelastic X-ray scattering, or IXS)~\cite{baron1}, and valence electrons (Raman scattering and resonant inelastic X-ray scattering, or RIXS)~\cite{ament2, devereaux2}. By probing the phonons through the lens of the valence electrons, one may also gain insight into the strength of the electron-phonon coupling relevant for the ground state and low energy excitations. Previous Raman spectroscopy measurements have observed both single phonon excitations and two-phonon excitations in the cuprates, including p-type La$_{2-x}$Sr$_x$CuO$_4$~\cite{sugai2}, YBa$_2$Cu$_3$O$_{6+\delta}$~\cite{chelwani1}, Bi$_2$Sr$_2$YCu$_2$O$_{8+\delta}$~\cite{chelwani1}, as well as n-type Nd$_{2-x}$Ce$_x$CuO$_4$~\cite{sugai1}. While the single phonon intensities remain generally similar with electron or hole doping with exceptions for some low energy modes, the two-phonon excitation intensity exhibit strong dependences on doping~\cite{sugai2}. Furthermore, while the two-phonon Raman intensity is strong, the three-phonon and four-phonon responses are not discernible~\cite{sugai2}. These early observations already suggest that the two-phonon excitation mechanism may be anomalous and beyond the simple high harmonic picture. 

Recent RIXS measurements found that a two-phonon excitation in Nd$_{2-x}$Ce$_x$CuO$_4$ appears to be hybridized with plasmons in the lightly doped and moderately doped regime~\cite{xu3}. This indicates that the two-phonon excitations may still strongly renormalize the low energy charge excitations in the doping range that is relevant to superconductivity and other emergent phenomena such as strange metallicity~\cite{martin1,chen1,greene1,mitrano1}. It is therefore imperative to examine the high order lattice excitations in the p-type cuprates and their undoped parent compounds. In this work, we study the prototypical p-type cuprate La$_{2-x}$Sr$_x$CuO$_4$ (LSCO) using RIXS at the O K-edge. Out of many prominent two-phonon peaks observed and identified in Raman spectroscopy~\cite{sugai2,chelwani1}, we observe only one two-phonon mode - the in-plane bond stretching mode - that couples significantly to the valence electrons. We also find a sharp peak at the energy of four bond-stretching phonons but notably the three-phonon excitation is missing. These high-order phonon excitations display strong dependence on hole doping. 

We first show the X-ray absorption spectroscopy (XAS) characterization of the samples measured in this study in order to establish which electronic bands are accessed at each incidence photon energy. Fig. 1(a) identifies the doping levels of crystals used for the RIXS study in the context of the LSCO doping phase diagram~\cite{croft1}. Fig. 1(b) shows the evolution of the O K-edge absorption, consistent with previous literature~\cite{ctchen1}. In the x = 0 sample, the lowest unoccupied band is the Cu $d_{x^2-y^2}$ upper Hubbard band (UHB), thus the first feature is the 531 eV feature corresponding to this band. As soon as some holes are doped into the system, the Fermi level jumps to the hybridized band formed between the Cu $d_{x^2-y^2}$ orbitals and O $p$ orbitals, also known as the Zhang-Rice singlet (ZRS)~\cite{zhang1}. With increasing doping, more holes are doped into the ZRS band and the intensity at 529.5 eV grows, while the intensity at the UHB decreases. For this study, we have focused only on the UHB as we are interested in the low doping levels where the ZRS intensity is very weak. Previous O K-edge RIXS studies on LSCO suggest that the charge excitations are similar between UHB and ZRS~\cite{nag2}.

With the identification of the incidence energy relevant to the UHB, we explore the doping dependence of the low energy excitations by performing RIXS. Fig. 2(a) shows the color plots of the energy loss features as a function of in-plane momentum $h$ along ($h$, 0 1), while Fig. 2(b) shows the raw spectra. Here, we have ignored the slight distortion of the crystal due to the low temperature orthorhombic structure~\cite{grande1,boni1}, and the momentum directions in this work refer to the tetragonal unit cell. The RIXS measurements were performed at a fixed $l$ value because the collective charge excitations may exhibit non-zero out of plane dispersion~\cite{hepting1}. In the x = 0 sample, there are at least 3 sharp peaks in the low energy range, in addition to a broad intensity around 400 to 600 meV ($\delta$). All of these excitations have no discernible momentum dependence at zero doping. At 1\% hole doping, the $\beta$ peak remains but the $\alpha$ and $\gamma$ modes are not clearly observable, likely due to broadening. Upon further doping to 5\%, the $\beta$ feature also broadens significantly, while the broad $\delta$ excitation is weakened. At 8\% doping, a dispersive feature ($\epsilon$) appears. At the same time, the non-dispersive broadened $\beta$ feature is still resolvable. The $\epsilon$ feature and $\beta$ features appear to cross each other. Within the crossing momentum around $h$ = 0.05 to $h$ = 0.07, the structure of the crossing is not clearly resolvable.  At the same time, the broad intensity around 500 meV ($\delta$ feature) is not resolvable. At 10\% doping, the only prominent feature (apart from the low energy single phonon below 50 meV) that remains is the dispersive $\epsilon$ excitation, although the existence of additional weak modes cannot be completely ruled out. The dispersive $\epsilon$ feature at x = 0.1 is consistent with the expectation of an acoustic plasmon, and the dispersion is comparable to the plasmon dispersion in an x = 0.16 sample from Ref.~\cite{nag2} (see supplemental materials Fig. 1).

Given the lack of resolvable dispersion in the RIXS spectra of the x = 0 samples, we can examine the energy scale of the modes by integrating the curves along different momenta. Fig. 3 shows the momentum-integrated RIXS energy distribution curve for the x = 0 sample compared with the Raman spectroscopy data from ref.~\cite{chelwani1}. The mode energies were found by fitting, and we obtain 85$\pm$5 meV for $\alpha$, 180$\pm$7 meV for $\beta$, 330$\pm$5 meV for $\gamma$. The broad $\delta$ feature is seen to be comprised of 2 features with $\delta_1 $ = 410 meV and $\delta_2$ = 560 meV. In compariosn, the Raman spectra of LCO also shows many excitations distributed across the $A_{1g}$ and $B_{1g}$ symmetry channels, which highlight different scattering processes. The low energy excitations below 200 meV in $A_{1g}$ are attributed to one-phonon and two-phonon scattering~\cite{sugai2,chelwani1}. We note that LSCO exhibits many more two-phonon excitations in the Raman spectra compared to other families of cuprates~\cite{chelwani1}. This is due to the fact that LSCO hosts a slightly orthorhombic crystal structure in the low temperature orthorhombic phase, such that additional crystal symmetries are broken compared to the usual tetragonal crystal structure of other cuprates. The broad peak near 400 meV in the $B_{1g}$ is putatively assigned to two-magnon scattering~\cite{lyons1}, although the origin of the shoulder feature near 550 meV is contentious~\cite{chelwani1,weidinger1}. Comparing the RIXS spectrum and the Raman spectra, we find that peaks $\alpha$ and $\beta$ in RIXS have corresponding features in Raman, albeit with larger widths due to the lower energy resolution of the RIXS measurement. On the other hand, not all the two-phonon peaks in Raman show up in RIXS. Remarkably, the sharp $\gamma$ feature seen in RIXS is not observed in Raman. The $\delta_1$ and $\delta_2$ features appear to have the same energy as the bimagnon feature and the post-peak shoulder feature in the $B_{1g}$ channel.

\begin{table}
\centering
\begin{ruledtabular}
\vspace{0.3cm}
\begin{tabular}{ccc}
\hline
\rule{0pt}{3ex} Peak label & Energy (meV) & \multicolumn{1}{c}{Putative assignment} \\
\hline
\vspace{0.1cm}
\rule{0pt}{3ex}$\alpha$ & 85$\pm$5 & \makecell{one-phonon bond stretching} \\ 
\vspace{0.1cm}
$\beta$ & 180$\pm$7  & two-phonon bond stretching  \\ 
\vspace{0.1cm}
$\gamma$ & 330$\pm$5 & four-phonon bond stretching  \\
\vspace{0.1cm}
$\delta_1$ &   410$\pm$26 &  bimagnon  \\
\vspace{0.1cm}
$\delta_2$ &    560$\pm$38  &  unknown  \\
\vspace{0.1cm}
$\epsilon$ &    dispersive &  plasmon  \\
\hline
\end{tabular}
\end{ruledtabular}
\caption{\label{tab:sample}Summary of RIXS excitations at the O K-edge for La$_2$CuO$_4$ and their putative assignments.}
\label{table1}
\end{table}

Having characterized the excitation energy scales and their evolution as a function of doping, we can now discuss the implications of the experimental observations and the possible microscopic origins. We first focus on the low energy sharp peaks in RIXS that likely relate to lattice excitations. Peak $\alpha$ at around 85 meV has the same energy as the single-phonon excitation involving the in-plane bond stretching mode identified in Raman measurements, while peak $\beta$ has approximately the same energy as the two-phonon excitation of the same mode. Thus, we putatively assign these two peaks to the same attribution as the Raman results. On the other hand, peak $\gamma$ at around 330 meV has no corresponding peaks in the Raman spectra in either $A_{1g}$ or $B_{1g}$ symmetry channels. Peak $\gamma$ cannot be a single-spin-flip magnetic excitation, as it has no dispersion and the 1$s$ core hole at the O $K$-edge resonance does not couple to the spin degree of freedom in the absence of strong spin-orbit coupling. Peak $\gamma$ also cannot be the bimagnon, as its linewidth is too sharp and its energy is incorrect. Remarkably, the $\gamma$ energy of around 330 meV is close to being four times the energy of the in-plane bond stretching phonon. Thus, in the absence of other known sharp excitations that can explain peak $\gamma$, we putatively assign it to the four-phonon excitation of this mode. 

At higher energies, we find that the $\delta_1$ feature in RIXS correspond nicely to the putative bimagnon feature in the Raman spectrum in terms of the energy position and width. The $\delta_2$ feature is apparently at approximately the same energy as the higher energy shoulder feature in the Raman spectrum. The origin of this higher energy shoulder feature is debated, with possible explanations being other types of two-magnon spin excitations~\cite{chelwani1} or a two-Higgs excitation~\cite{weidinger1}. However, since the intensity of the higher energy shoulder feature seems to be strongly dependent across cuprate materials families~\cite{chelwani1}, this feature may not be integral to the salient physics in the cuprates. The summary of the RIXS excitations and their putative assignments are detailed in Table~\ref{table1}.

Compared to Raman scattering, which couples to many different valence bands within a few eV of the Fermi level depending on the excitation wavelength, RIXS is extraordinarily selective to both the chemical species and the orbital states. The energy selectivity is mainly limited by the core hole broadening, which is typically about 0.3 eV at the O K-edge~\cite{frati1}. Even though Raman scattering can resolve the presence of many two-phonon excitations enabled by symmetry-breaking in the orthorhombic phase, our results directly show that only the in-plane bond stretching mode is strongly coupled to the UHB states for the RIXS intermediate state. This is an important phenomenological conclusion regardless of the microscopic mechanisms inducing these high-order modes. 

Furthermore, the observation of putative four-phonon excitations is unexpected, especially considering the absence of any appreciable intensity from the three-phonon excitation. Remarkably, there are no excitations associated with this energy scale that can be resolved in Raman spectroscopy. While high order harmonics of phonon excitations can be observed in systems with extremely strong electron-phonon coupling~\cite{ament1, vale1, hong1, wslee2}, where the vibrational excitations have intensities following the Franck-Condon envelope, the phenomenology observed here is inconsistent with the conventional understanding. In LCO, the absence of the three-phonon mode suggests physics beyond this conventional picture. We note that the scenario of symmetry-enforced excitation of even order phonons likely does not apply here due to the orthorhombic crystal structure and absence of four-phonon excitations in Raman spectra.

One possible interpretation is that the even-ordered phonon excitations are associated with nonlinear interactions and bound bipolarons. Here, we use the term bound bipolaron to refer to a sharp excitation comprised of a pair of valence charger carriers coupled to a pair of lattice distortions trapped in a local potential well. Because the lattice around the pair is already displaced from its equilibrium configuration, the phonon dynamics are no longer well described by weakly interacting harmonic vibrations. Instead, the local distortion produces an effective nonlinearity and can mediate an attractive interaction between phonons, allowing two phonons to form a locally bound state~\cite{franchini1} rather than just appearing as an incoherent two-phonon continuum. Mathematically, the RIXS intermediate state is coupled to the lattice displacement with both linear and quadratic terms

\begin{equation}
    H_{int} = g_1 Q n_{int} + g_2 Q^2 n_{int} + \mathcal{O}(Q^3)
\end{equation}

where $n_{int}$ denotes the occupation of the core-excited electronic state. The linear term can generate the one-phonon excitation, while the $Q^2$ term can create or annihilate phonons in pairs. Therefore, a sufficiently strong quadratic coupling may enhance the two-phonon and four-phonon peaks while retaining relatively weak intensity for the three-phonon peak.

The non-linear coupling to the lattice and bound bipolarons may be also related to the idea that the underdoped and undoped cuprates have strong pairing potential~\cite{emery1} and potentially host localized Cooper pairs. While highly speculative, such localized pairs would be consistent with the bipolaronic picture, in which pairing and lattice distortion are tied together locally. In this case, the attraction provided by the nonlinear lattice interactions may operate synergistically with the magnetic degrees of freedom~\cite{anderson1}. However, global superconductivity fails to emerge because the pairs do not delocalize and establish long-range phase coherence.

Regardless of the microscopic origin of the high-order phonon modes, the unusual selective coupling and persistence of the multi-phonon excitation into the doping regime where superconductivity emerges at x $>$ 0.05 suggests that these high order phonon excitations may play an important role in shaping the low energy electronic landscape relevant for superconductivity. Further studies are required to elucidate the nature of these excitations and the relation to superconductivity, including a survey of different materials families and real space probes such as scanning tunneling microscopy to check for signature of localized pairs in the deeply underdoped regime.

\hspace{0.5cm}

\begin{acknowledgments}
We thank B. Moritz for insightful discussions. The work performed at the University of California, Berkeley and Lawrence Berkeley National Laboratory was funded by the U.S. Department of Energy, Office of Science, Office of Basic Energy Sciences, Materials Sciences and Engineering Division, Contract No. DE-AC02-05-CH11231 within the Quantum Materials Program (KC2202). The work performed at Stanford and SLAC was supported by the U.S. Department of Energy (DOE), Office of Basic Energy Sciences, Division of Materials Sciences and Engineering, Contract No. DE-AC02-76SF00515. J. C. and S. K. acknowledge support from the National Research Foundation of Korea (NRF) grant funded by the Korean government (MSIT) RS-2026-25491227 and support from Creative-Pioneering Researchers Program and New Faculty Startup Fund through Seoul National University. This work made use of Beamline 41A of the Taiwan Photon Source at the National Synchrotron Radiation Research Center (NSRRC), Hsinchu, Taiwan, under proposal number 2024-2-251.
\end{acknowledgments}

\hspace{0.5cm}

\bibliographystyle{apsrev4-2}
\bibliography{main_ref}
\raggedend
\end{document}